\documentclass{article}
\usepackage{subequation,graphicx}
\def\circa#1{\,\raise.3ex\hbox{$#1$\kern-.75em\lower1ex\hbox{$\sim$}}\,}
\def\bom#1{{\mbox{\boldmath $#1$}}} \newcommand{\ov}{{\cal O}}

\newcommand{\f}{{\cal F}} \newcommand \bra {\langle} \newcommand \ket
{\rangle} \newcommand{\be}{\begin{equation}}
\newcommand{\ee}{\end{equation}} \newcommand{\ben}{\begin{displaymath}}
\newcommand{\een}{\end{displaymath}} \newcommand{\ba}{\begin{eqnarray}}
\newcommand{\ea}{\end{eqnarray}} \newcommand{\ban}{\begin{eqnarray*}}
\newcommand{\ean}{\end{eqnarray*}} \newcommand{\cro}{\dagger}
 \newcommand{\de}{\partial}

\newcommand{\kv}{{\bf k}}   \newcommand{\Ttot}{{\bf T}} 
 
\newcommand{\tv}{{\bf t}}
\newcommand{\Tv}{{\bom T}}

 \catcode`\@=11

\begin{document}

\vspace{1.cm}

{\centering
      
{\Large\bf Electroweak double-logs at small $x$
}

\vspace{1.cm}

{\bf \large Marcello Ciafaloni}

{\it Dipartimento di Fisica, Universit\`a di Firenze and INFN - Sezione di
 Firenze,\\ via Sansone 1, I-50019 Sesto Fiorentino, Firenze, Italy\\
 E-mail: ciafaloni@fi.infn.it }

\vspace{0.4cm}

{\bf \large Paolo Ciafaloni}

{\it Dipartimento di Fisica, Universit\`a di Lecce and INFN - Sezione di
Lecce, \\ Via per Arnesano, I-73100 Lecce, Italy 
\\ E-mail: paolo.ciafaloni@le.infn.it}
\vspace{0.4cm}

{\bf \large Denis Comelli}

{\it INFN - Sezione di Ferrara, \\Via Saragat 3, I-44100 Ferrara, Italy\\
E-mail: comelli@fe.infn.it}

}

\vspace{0.3cm}

\begin{abstract}
  We investigate enhanced EW corrections to inclusive hard processes in the
 TeV energy region with emphasis on the small-$x$ situation, in which the
 hard scale $Q$ is significantly smaller than the available energy
$\sqrt{s}\equiv Q/x$. We
 first propose and justify a general factorization formula in which
the (double-log) EW form factor at scale $Q^2$ is factorized from
EW parton distribution
 functions, which satisfy evolution equations of DGLAP
type.
We then investigate the small-$x$ behavior of the EW parton distributions
including the novel ones for non-vanishing t-channel weak isospin T 
 and we compare it with a
 BFKL-type approach.
In either approach we find that large small-$x$
 corrections of order $ \alpha_w \,\log x\, \log Q^2/M^2 $ ($M$ being the EW
 symmetry breaking scale) are present only for T=2 and not for T=1.
This implies that
only transverse $WW$ interactions (coupled to T=2) are affected, while the T=1
components feel just the form factor at scale $Q^2$.
\end{abstract}

\section{Introduction}

It is by now clear that electroweak radiative corrections at the TeV
scale~\cite{sirlin, CC} have a much richer structure and a higher
phenomenological relevance than one could have thought, say, ten years ago.
The size of the corrections, for one thing, is typically of the order of
20-30 \%, much bigger than the LEP {\sl permille} level. This is due to the
fact that the EW (Electro-Weak) corrections grow like the log square of the
c.m. energy, which in turn is tied to the infrared structure of the theory
\cite{CC} and opens up the possibility of resumming leading effects, with
techniques partly mutated from QCD \cite{FadinCC}.

Even more interestingly, the infrared structure of the electroweak sector
is radically different from QED or QCD, due to spontaneous symmetry
breaking. As a result, the double log dependence on the infrared cutoff,
which is physical and of the order of the weak scale, is present in both
exclusive {\sl and} fully inclusive observables\footnote{The ``inclusive''
qualification refers, as usual, to observables which sum over all
unobserved particles, including final state unobserved weak gauge bosons.},
a phenomenon baptized ``Bloch-Nordsieck violation''\cite{CCC00}. This is a
very interesting and striking fact from a theoretical point of view;
phenomenologically, it means that considering the possibility of weak W,Z
gauge bosons emission is more important than one could think. In fact, in
the case of ILC physics EW corrections can dominate over the QCD ones
\cite{CCC00}, and even for an environment which is {\sl a priori} dominated
by QCD, like the LHC, studying weak boson emission turns out to be
important \cite{CC06,Baur}.

In the above framework, various kinds of electroweak corrections for
TeV-scale observables have been considered: exclusive observables -- with
an extensive literature on one loop results \cite{excl} and two loop
calculations \cite{2loop} -- and inclusive observables \cite{CCC01},
featuring the noncancellation phenomenon.  In the latter case, the hard
subprocess scale $Q^2$ has always been assumed to be of the same order as
the initial c.m. energy $s$.  This is the case for $e^+e^-\to jj X$ where
$j$ represents a final jet and $X$ includes gauge bosons radiation, when
$s\circa{>} Q^2\gg \mu^2$, $Q^2$ being the 2 final jets invariant mass and
$\mu$ the infrared cutoff scale, of the order of the weak scale.

The purpose of this work is to investigate the behavior of high energy
electroweak corrections to fully inclusive observables when the hard
subprocess scale $Q^2$ is significantly different from the c.m. initial
scale $s$, i.e. when $\frac{Q^2}{s}$ becomes small.

The above problem is relevant and interesting under several
aspects. Firstly, from a phenomenological point of view many processes
relevant for both LHC and ILC physics (like Z and W production, Higgs
production and so on) are characterized by relatively small values of
$x$. Secondly, from a theoretical point of view the issue of factorization
of electroweak corrections, because of the presence of uncanceled double
logs and three different scales $Q^2, s, \mu^2$ is far for being
trivial. On top of this, the presence of big corrections related to the
collider scale $s$ of the form $\log\frac{s}{\mu^2}$ would mean that
electroweak corrections receive huge enhancements even for small $Q^2$
values: addressing this issue is of course of paramount importance.  To end
with, as we shall see, small-$x$ electroweak physics is qualitatively
different from QCD; namely the relationship between DGLAP \cite{DGLAP} and
BKFL \cite{BFKL} approaches has a richer structure because of the
uncanceled double-logs in inclusive observables.

\section{Flavour structure of inclusive electroweak double-logs}
Let us first recall the flavour structure\footnote{by ``flavour'' here we
mean ``weak isospin or hypercharge quantum numbers''. We only consider one
family, and therefore neglect family mixing.}  of inclusive electroweak
corrections which are infrared sensitive, and thus contain, in the TeV
energy region, double logarithms of the symmetry breaking scale.  The
effective coupling for such corrections is \be
\alpha_{eff}(Q^2)~\equiv~\alpha_W\log^2(Q/M_W)~~~~(\alpha_w\equiv
g_w^2/4\pi) \ee which is of the order $\alpha_{eff}=0.2$ at the TeV
threshold.  The treatment of such corrections has been performed so far at
both double-log and single-log accuracy, by proposing evolution equations
\cite{CCC01} which generalize the DGLAP equations \cite{DGLAP} to a
non-trivial flavour structure. Our final purpose is to investigate such
equations at small Bjorken-$x$, and to compare with the analogue BFKL
evolution equation \cite{BFKLew}. Before doing that, it is useful to
understand the flavour dependence of such corrections in the eikonal limit
-- which takes into account double logs while neglecting single logs -- and
then to look at the collinear factorization structure for $x\equiv Q^2/s\ll
1$.

We shall first consider inclusive, flavour-blind observables, triggered by
initial fermions or bosons. Typical examples, for ILC physics would be just
$e^+~e^- \rightarrow $ hadronic jets or, for LHC physics,
$q~\bar{q}\rightarrow$ jets or $W^+~W^-\rightarrow$ jets. In such cases, no
flavour is registered in the final state, while the initial states carry
their own, and we shall focus on their momenta and weak isospin, denoted by
($p_1,\alpha_1;p_2,\alpha_2 $) for the two incoming particles. These hard
processes are therefore typically of Drell-Yan type. Due to the inclusive
nature of the cross-sections so defined, final state singularities cancel
out \cite{CCC00}, and the remaining dependence is on initial state isospin
indices, so that the flavour dependence of the squared amplitude can be
arranged in an overlap matrix with two initial and two ``final'' indices,
which actually double the initial ones. We shall comment later on the
possible generalization to registered flavour in the final state, a case in
which the overlap matrix involves more indices.

\begin{figure}
      \centering \includegraphics[height=45mm] {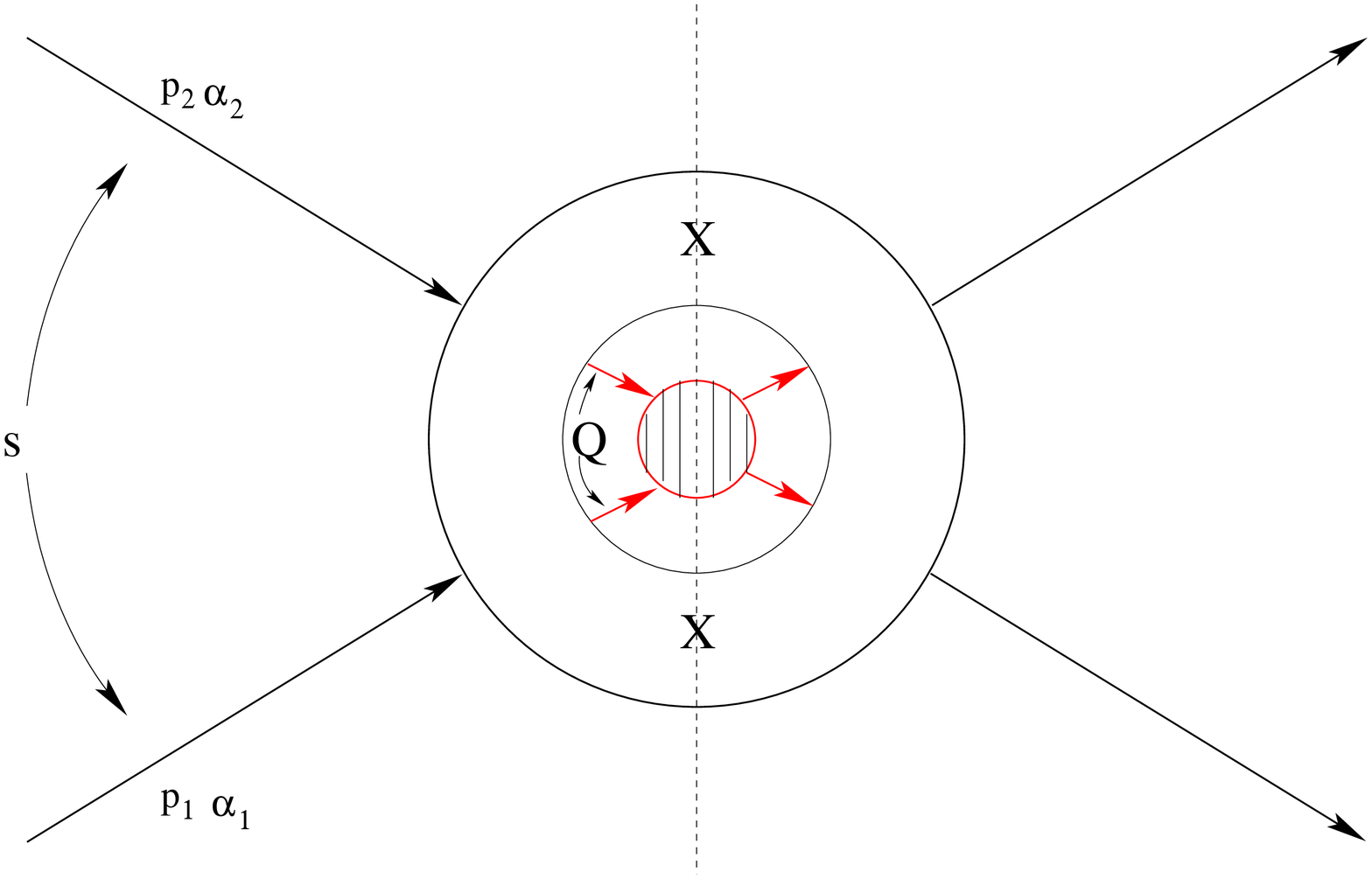}
      \caption{\label{fig1} $2 \leftrightarrow 2$ overlap matrix for a
 Drell-Yan-type process, with two initial states of flavour $\alpha_{1,2}$
 and momenta $p_{1,2}$ with $(p_1+p_2)^2=s$. In the center (in red), the
 observed flavour-blind process at hard scale $Q$.  Inside $X$ we
 schematically include the undetected background (from QCD, QED and
 possibly IR and collinear W emissions ).  }
         \end{figure}

To be more precise, in order to describe the isospin dependence of EW
 corrections, one can define the overlap matrix illustrated in Fig.1, which
 generalizes the cross-section for the flavour dependence, and is given in
 terms of the T-matrix as follows \cite{CCC00} ($\alpha_i,\beta_i$ are the
 initial state isospin indices): \be\label{overlap} \bra p_1 \beta_1,p_2
 \beta_2|\;T^\cro \;T\;|p_1\alpha_1,p_2\alpha_2\ket
 =\;\ov_{\beta_1\alpha_1,\beta_2\alpha_2}^{p_1p_2} \ee which can be roughly
 related by the optical theorem to the imaginary part of the forward
 amplitude. The observable cross sections are then obtained for diagonal
 flavour indices as follows: \be
 d\sigma_{\alpha_1\alpha_2}=\ov_{\alpha_1\alpha_1,
 \alpha_2\alpha_2}^{p_1p_2}~d\Phi \ee where we define the overlap matrix to
 be dimensionless, so that the phase space is meant to be rescaled by a
 proper power of $s$.  The SU(2) generators $t^a_i,t'^a_i, a=1,2,3\quad
 i=1,2$ act on the overlap matrix as in the following example \be (t^a_1\ov
 )_{\beta_1\alpha_1,\beta_2\alpha_2}=
 \sum_{\alpha'_1}(-t^a_{\alpha'_1\alpha_1})
 \ov_{\beta_1\alpha'_1,\beta_2\alpha_2} \qquad (t'^a_1
 \ov)_{\beta_1\alpha_1,\beta_2\alpha_2}=
 \sum_{\beta'_1}t^a_{\beta_1\beta'_1}
 \ov_{\beta'_1\alpha_1,\beta_2\alpha_2} \ee and are of course dependent on
 the representation of the considered $i$-th particle.

Virtual and real emission of soft gauge bosons in the initial state for the
overlap function (\ref{overlap}) in Fig.1 is provided, at leading
double-log level, by the external line insertion of the eikonal current \be
\label{eikcur}
{\bom J}^{\mu }(k)=g_w\;\sum_{i=1}^{2} {\bom
T}_i\frac{p_{i}^{\mu}}{p_i\cdot k} ={\bom T}_1 (\frac{p_{1}^{\mu}}{p_1\cdot
k}- \frac{p_{2}^{\mu}}{p_2\cdot k})\equiv\Tv_1j_{12}^{\mu}(k)\, \ee where
$k$ is the momentum of the emitted soft gauge boson, $p_i$ the i-th leg
momentum, $g_w$ the SU(2) gauge coupling and we have defined the total
($t$-channel) isospin generator referred to the leg $i$ as $T_i\equiv
t_i+t'_i$ \cite{CCC00}.  Notice that the part of the current proportional
to $g'$ is absent altogether because of the cancellation of the abelian
components for inclusive observables \cite{CCC00}.  Note, furthermore, that
we have used in (\ref{eikcur}) isospin conservation to set $\Tv_2=-\Tv_1$
. In fact, since we consider energy scales of the order of 1 TeV and
beyond, we take all particles to be massless and we work in the limit in
which the $SU(2)\otimes U(1)$ symmetry is recovered. In other words the
overlap matrix is invariant under total isospin transformation:
\be\label{sym} {T}_{tot}^a\equiv\sum_i { T}_i^a\qquad \exp[\alpha^a {
T}^a_{tot}]\ov= \exp[\bom{\alpha}\cdot \bom{ T}_{tot}]\ov=\ov\quad
\Rightarrow \quad \bom{T}_{tot}\ov=0 \ee

The emission probability of real and virtual bosons off the initial legs is
 then obtained by squaring the eikonal current so as to obtain the
 insertion operator\be\label{insertion} \frac12 {\bom J}_{\mu}{\bom
 J}^{\mu}=-g_w^2\frac{p_1p_2}{(kp_1)(kp_2)}{\bom {T}}_1^2;~~~-\Tv_1^2=
 -\tv_1^2-{\tv'}_1^2-2\tv_1\cdot {\tv'}_1=-2C_1-2\tv_1\cdot{\tv'}_1 \ee
 which in turn provides the eikonal radiation factor for gauge boson
 emission: \be\label{radiator}
 L_W(s)=\frac{g_w^2}{2}\int_M^E\frac{d^3\bom{k}}{2\omega_k(2\pi)^3}
 \frac{2p_1p_2}{(kp_1)(kp_2)}=\frac{\alpha_w}{4\pi}\log^2\frac{s}{M^2},
 \quad\alpha_w=\frac{g_w^2}{4\pi} \ee where the $k$-integral has been
 performed over the soft momentum fraction region \be\label{cutoff}
 (1-z)\sqrt s > |\kv| > M_Z\simeq M_W \equiv M \ee 
in both the forward 
and backward 
hemispheres with
 respect to the incoming momentum $p_1$. 
Finally, by iterating the procedure over any number of
 soft bosons, we get the resummed expression for the overlap matrix as 
 the simple exponential \be\label{dress} \ov(s)=\exp[L_W(s)
 \bom{T}_1\cdot\bom{T}_2]\ov^H=\exp[-L_W(s)(\Tv_1^2+\Tv_2^2)/2]=\exp\left[-
 L_W(s) \bom{T}_1^2\right]\ov^H \ee

Note that, due to the simple relation in eq.~(\ref{insertion}) of the total
$t$-channel isospin to the Casimir operators of the colliding particles,
real emission ($\sim -2\tv_1\cdot {\tv'}_1$) occurs with weight
$(2C_1-\Tv^2)$, relative to the virtual corrections provided by $2C_1$.
This structure will be essentially kept at single-log level as well, in the
DGLAP and BFKL approaches that we shall consider next for $x\equiv Q^2/s\ll
1$.

\section{Collinear factorization and EW evolution equations}

\begin{figure}
      \centering \includegraphics[height=35mm] {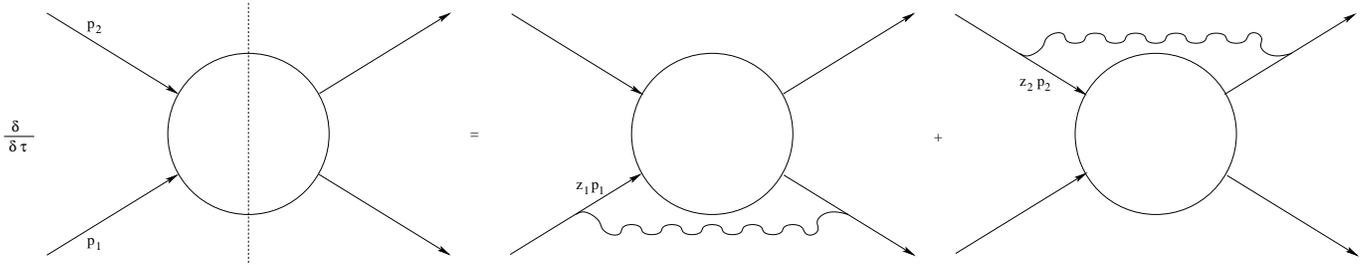}
      \caption{\label{fig2} General structure of the Collinear Evolution
Equations for the $2 \leftrightarrow 2$ overlap matrix
.  The wavy lines denote real boson emission with momentum rescaling.  }
         \end{figure}

Let us now recall the collinear evolution equations for the overlap matrix,
with the purpose of resumming single and double EW logarithms. Such
equations were derived in \cite{CCC01}, where, with the help of collinear
Ward Identities and working in the Feynman gauge, it was shown that the
only relevant diagrams are those illustrated in Fig. 2.

 Here, in order to discuss the small-$x$ limit, we specialize to the case
of initial bosons and we shall simplify it by omitting the mixing with the
fermionic channels.  Then the above procedure results in the following
bosonic channel evolution equation for the overlap matrix at given initial
energy-squared $s$, measurement hard scale $Q^2$ and t-channel isospin
$\Tv^2=T(T+1)$: \ba\label{ov_ev} \frac{\de \ov^{(T)}(s,Q^2,\mu^2)}{\de
\tau}& =&\frac{\alpha_W}{\pi}\left\{
2\, p^V\;\frac{\Ttot^2}{2}\;\ov^{(T)}(s,Q^2,\mu^2)\right.  \\\nonumber &+&\left.
(C_A-\frac{\Ttot^2}{2})\int \frac{dz_1}{z_1}
P(z_1)\;\ov^{(T)}(sz_1,Q^2,\mu^2) +(C_A-\frac{\Ttot^2}{2})\int
\frac{dz_2}{z_2} P(z_2)\;\ov^{(T)}(sz_2,Q^2,\mu^2) \right\} \ea \ba
P^V&=&p^V\delta(1-z),~~~p^V=-\left(\frac{1}{2}\log\frac{s}{\mu^2}
-\frac{11}{12}\right)\qquad\\ \nonumber
P^R&=&\left(z(1-z)+\frac{z}{1-z}+\frac{1-z}{z}\right)
\theta(1-z-\frac{\mu}{\sqrt{s}});~~~ P(z)=\lim_{\frac{\mu}{\sqrt{s}}\to
0}(P^V+P^R) \ea where $\tau\equiv\log(Q^2/\mu^2)$ is the evolution
parameter, $\mu^2<\kv^2$ is an infrared cutoff on transverse momenta (to be
set equal to the symmetry breaking scale at the end) and $z_1$ ($z_2$) is
the momentum fraction variable for DGLAP splitting on leg $1$ ($2$).  Note
that the $P^V$ distribution is cutoff dependent, the real emission density
$P^R$ incorporates the soft emission cutoff in (\ref{cutoff}), while $P(z)$
is the regularized one, obtained by combining $P^R$ and $P^V$.
The
normalization of $P(z)$ differs by a factor of $2$ from the customary one,
so that $P(z)\sim 1/z$ for $z\rightarrow 0$.

The overlap evolution equation (\ref{ov_ev}) was already used in
 \cite{CCC01} to introduce bosonic (or fermionic) PDFs and to derive
 DGLAP-type equations for them in the case in which $s$ and $Q^2$ are of
 the same order. Here we want to generalize this procedure to the small-$x$
 region, where the collinear factorization has to specify which scale,
 $Q^2$ or $s$, has to carry the EW double-logs in order to be able to
 factorize the appropriate PDFs. We shall show that $Q^2$ is the
 appropriate choice and, to this purpose, we propose the following
 factorization ansatz: \be\label{fact}
 \ov^{(T)}(s,Q^2,\mu^2)=\int\frac{dx_1}{x_1}\frac{dx_2}{x_2}\;
 \exp\left[-\frac{\alpha_w}{2\pi}\frac{\Ttot^2}{2}(\log\frac{Q^2}{\mu^2})^2\right]
\;\; f^{(T)}(x_1,Q^2,\mu^2)\;\;\ov^{(T)}_H(\frac{Q^2}{x_1,x_2s})\;\;f^{(T)}(x_2,Q^2,\mu^2)
 \ee where the bosonic PDFs $f_i^{(T)}\equiv f^{(T)}(x_i,Q^2,\mu^2)$)
 (normalized so that $xf_i^{(T)}$ are quasi-constant in the small-$x$
 region) are supposed to be free of double-logs in $Q^2/M^2$ but still
 contain $\log Q^2/\mu^2$ and $\log 1/x$ enhancements, while $\ov_H$ should
 not contain any collinear or high-energy logarithms. The above
 factorization property is a non-trivial extension of the one valid in QCD,
 because -- for $\Ttot^2\ne0$ -- it is meant to control both EW double-logs
 and single collinear logs. 

In order to derive the factorization ansatz (\ref{fact}) at collinear
 level, we replace it in (\ref{ov_ev}) by neglecting for simplicity the
 single-logs generated by  $p^V$ (they can be restored later on) because 
we are mostly
 interested in controlling double-logs of IR-collinear type and mixed ones,
 involving high-energy logarithms $\log 1/x$. We then obtain an evolution
 equation involving the product of PDFs: \be\label{f1f2} \frac{\de}{\de
 \tau}\left[f_1^{(T)}f_2^{(T)}\right]
 =-\frac{\alpha_w}{\pi}\frac{\Ttot^2}{2}
 \left(\log\frac{1}{x_1}+\log\frac{1}{x_2}\right)f_1^{(T)}f_2^{(T)}+
 \frac{\alpha_w}{\pi}\,(C_A-\frac{\Ttot^2}{2})\;\left[(P\otimes
 f_1^{(T)})f_2^{(T)}+ f_1^{(T)}(P\otimes f_2^{(T)})\right] \ee where
 $P\otimes f(x)\equiv\int \frac{dz}{z}P(z)f(\frac{x}{z})$, and we note that
 the relation $s=Q^2/x_1x_2$ has generated the additive virtual term $\sim
 \log(x_1x_2)$.  Finally, by dividing by $f_1f_2$ we obtain the single leg
 evolution equation: \be\label{eqf} \frac{\de f^{(T)}(x,\tau)}{\de\tau}=
 -\frac{\alpha_w}{\pi}\,\frac{\Ttot^2}{2}\,\log\frac{1}{x}\;f^{(T)}(x,\tau)
 +\frac{\alpha_w}{\pi}\,(C_A-\frac{\Ttot^2}{2})\;P\otimes
 f^{(T)}(x,\tau),~~~(0\le\tau\le t\equiv\log\frac{Q^2}{M^2}) \ee which
 therefore exhibits the desired factorization.  We remark the role of
 choosing the $Q^2$ scale for the double-log factor in (\ref{fact}), in the
 above derivation of factorized evolution equations.  Had we chosen the
 energy variable $s=Q^2/{x_1x_2}$, it would have generated nonfactorizable
 double logs of type $\log x_1\log x_2$ thus violating the factorized
 structure of (\ref{f1f2}).

We can also investigate inclusive hard processes in which some flavour
 (weak isospin) state is identified in the final state for one or more
 particles, similarly to multiparticle distributions in QCD. The
 corresponding overlap function is now an $n$ by $n$ matrix, as depicted in
 Fig. 3, where we denote initial $t$-channel isospins by $\Tv_1$ and
 $\Tv_2$ and final ones by $\Tv_3, \dots, \Tv_n$.
We believe that a factorization formula exists in this case also, by
\begin{figure}
      \centering \includegraphics[height=45mm] {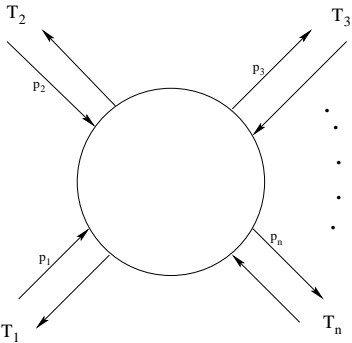}
      \caption{\label{fig3} Overlap matrix with $n$ registered flavour
isospin legs (they can be both in the initial that in the final state) }
\end{figure}
singling out an $n$-particle form-factor computed by the eikonal current
 insertion as in the $2 \rightarrow 2$ case, while the collinear logarithms
 are factored out in PDFs or final state fragmentation functions.  The
 situation is here similar to that occurring in QCD when the phase-space
 boundary or some veto expose a double-log behaviour in infrared sensitive
 observables~\cite{BCM}.  As such, this problem has been investigated since
 quite a time~\cite{SCB} and, most recently, in~\cite{DM}.  The eikonal
 squared current is well known for $n=3$.  By using isospin conservation
 ($\Tv_1+\Tv_2+\Tv_3=0$) we can generalize eqs. (\ref{eikcur}) and
 (\ref{insertion}) to obtain \be \frac12{\bom J}_{\mu}{\bom
 J}^{\mu}=\frac12\; \Tv_1^2\; j_{12}(k)\cdot j_{13}(k)~+~{\rm cyclic}
 ~~~~~(n=3) \ee where the product of currents in front of $\Tv_i^2$, by the
 definition in (\ref{eikcur}), is collinear singular only for $\kv$ in the
 direction of $p_i$.  Therefore, at double-log level, the eikonal radiation
 exponent is simply additive in the isospin charges, as follows \be
 \ov^{(T)}_n\sim\exp({\cal L}_W^{12\cdots n}),~~~{\cal L}_W^{12\cdots
 n}=-\frac12 \;L_W(Q^2)\;\sum_{i=1}^n \Tv_i^2 + {\rm single~ logs} \ee A
 similar result holds for $n=4$, except that the single IR logarithms
 should now be computed by the techniques described in~\cite{SCB, DM}.

The novel feature of the evolution equation (\ref{eqf}) for the PDFs--
 compared to the DGLAP equations in QCD -- is the $\log 1/x$ behaviour of
 the diagonal term for $\Tv^2\ne 0$, which suggests the existence of extra
 {\it damping} due to mixed infrared and high-energy logarithms ($\tau\;log
 \frac{1}{x}$), to be looked at in detail. In fact, in this
 nontrivial-flavour evolution, such suppression could modify in an
 important way the basic form-factor behaviour already factored out in
 (\ref{fact}) (the factor
 $\exp[-\frac{\alpha_w}{2\pi}\frac{\Ttot^2}{2}(\log\frac{Q^2}{\mu^2})^2]$). We
 shall basically investigate that in the following, by using both DGLAP and
 BFKL approaches.  We further notice that the flavour factors occurring in
 (\ref{eqf}) for the form-factor vs. real-emission terms are indeed the
 same as in the eikonal treatment recalled before, where the relevant
 Casimir is now that of the adjoint representation, $C_A=2$. The possible
 values of $\Tv^2=T(T+1)$ are instead provided by $T=0,1,2$ as usual.

\section{Bosonic DGLAP-type equation in the small-$x$ region}

Here we shall solve by customary methods the bosonic equation (\ref{eqf})
for the various $T$ values, by focusing on its small-$x$ behaviour.  By
introducing the Mellin transform variable $\omega\equiv N-1$, where $N$ is
the customary moment index 
\be \tilde{f}(\omega)=\int_0^1 dx \;x^\omega\;
f(x)\;,\qquad x f(x)=\frac{1}{2\pi i}\,\int_{c-i\infty}^{c+i\infty}
d\omega\; \tilde{f}(\omega)\;x^{-\omega} \ee 
the basic equation (\ref{eqf}) can be
rewritten in differential form 
\be\label{om_ev} \frac{\de
\tilde{f}^{(T)}(\omega,\tau)}{\de\tau}= 
\frac{\alpha_w}{\pi}\;\frac{\Ttot^2}{2}\;
\frac{\de \tilde{f}^{(T)}(\omega,\tau)}{\de\omega}
+\frac{\alpha_w}{\pi}\;\left(C_A-\frac{\Ttot^2}{2}\right)\;\tilde{P}(\omega)\;
\tilde{f}^{(T)}(\omega,\tau),~~~\tilde P(\omega)\equiv\int_0^1dz~z^{\omega}P(z) 
\ee 
The
solution simplifies in the cases $T=0,1$ that we shall consider separately.

\subsection {Solutions for T=0 and T=1}

 To start with, for $T=0$, (\ref{om_ev}) takes a DGLAP form. In the
small-$x$ limit, with $P(z)= 1/z\Rightarrow P(\omega)=1/\omega$, given the
initial conditions $\tilde{f}^{(0)}(\omega,\tau=0)\equiv 
\tilde{f}_0^{(0)}(\omega)$, we
obtain 
\be
\tilde{f}^{(0)}(\omega,\tau)=\tilde{f}_0^{(0)}(\omega)
\;\exp\left[\frac{\alpha_W}{\pi}
\frac{C_A}{\omega}\tau \right] 
\ee 
and, by antitrasforming: 
\be\label{T_0} x
f^{(0)}(x,\tau)=\frac{1}{2\pi i}\int_{c-i\infty}^{c+i\infty}d\omega\;
\exp\left[\omega\;\log\frac{1}{x}+\frac{\alpha_W}{\pi}
\;\frac{C_A}{\omega}\;\tau\right]\;\;\tilde{f}_0^{(0)}(\omega),~~~(T=0)
\ee 
When $\log\frac{1}{x}\gg 1$ this integral can be evaluated by a saddle
point method and we get the usual double-log DGLAP behaviour, corresponding
to a cross-section increase in the small-$x$ region: \be\label{DL} x
f^{(0)}(x,\tau)\simeq \left(\frac{\alpha_w C_A\tau/\pi\log\frac{1}{x}}{4\pi
\sqrt{\alpha_w C_A\tau\log\frac{1}{x}}}\right)^{1/2}\exp\left[2\sqrt{\frac{\alpha_w}{\pi}C_A\tau\log\frac{1}{x}}\right]
\quad
\tilde f_0^{(0)}\left(\omega=\sqrt{\frac{\alpha_wC_A\tau}{\pi\log\frac{1}{x}}}\right),~~~(T=0)
\ee

The case $T=1$ is interesting too, but shows a quite different behaviour
 whose general form will be found below. Here we note a special small-$x$
 solution of eq.~(\ref{eqf}) in the case $P(z)\simeq 1/z$, provided simply
 by \be\label{T_1} x f^{(1)}(x,\tau)~=~F_1=const.~~~~(T=1) \ee The reason
 for such a constant solution is that the flavour factors for virtual and
 real emission terms become equal for $T=1$, namely
 $\Tv^2/2=C_A-\Tv^2/2=1$.  Therefore, the $\log 1/x$ evolution factor
 cancels out between virtual and real emission contributions. A subleading
 $x,\tau$ dependence survives, and is found below for given boundary
 conditions. However the fact remains that, for $T=1$, there are no
 double-log corrections ($\tau\;log\frac{1}{x}$) to the basic form factor
 factorized in (\ref{fact}).

\subsection{Solutions for generic $\Ttot$ values}

For generic values of $T$, we can integrate eq.~(\ref{om_ev}) to get the
general solution 
\be
\tilde{f}^{(T)}(\omega,\tau)=\Phi(\frac{\alpha_w\Ttot^2}{2\pi}\;\tau+\omega)
\;\;\exp\left[-(\frac{2C_A}{\Ttot^2}-1)\;\;\int_c^\omega d\omega'
\;\;\tilde{P}(\omega')\right] \ee 
$c$ being an arbitrary constant and $\Phi$ an
arbitrary function, to be determined through initial conditions. If we
demand that, at $\tau=0$, $\tilde{f}^{(T)}(\omega,\tau=0)=\tilde{f}_0^{(T)}(\omega)$ then,
since $C_A=2$: \be
\tilde{f}^{(T)}(\omega,\tau)=\,\tilde{f}_0^{(T)}
(\omega+\frac{\alpha_w\Ttot^2}{2\pi}\tau)\quad
\exp\left[(\frac{4}{\Ttot^2}-1)\;\int_\omega^{\omega+
\frac{\alpha_w\Ttot^2}{2\pi}\tau}d\omega' \tilde{P}(\omega') \right] 
\ee 
and, by
antitrasforming to $x$ space, we find that an extra form factor is
factorized out as follows: \ba x f^{(T)}(x,\tau)&=
\int_{c-i\infty}^{c+i\infty}\frac{d\omega}{2\pi i}\;x^{-\omega}\;
\tilde{f}_0^{(T)}(\omega+ \frac{\alpha_w\Ttot^2}{2\pi}\tau)\;\;\exp
\left[(\frac{4}{\Ttot^2}-1)\;\int_\omega^{\omega+
\frac{\alpha_w\Ttot^2}{2\pi}\tau}d\omega' \tilde P(\omega') \right ] \\
&=e^{-\frac{\alpha_w\Ttot^2}{2\pi}\tau\log \frac{1}{x}}\;\;
\int_{c-i\infty}^{c+i\infty}\frac{d\omega}{2\pi i} \;x^{-\omega}\;
\tilde{f}_0^{(T)}(\omega)\;\;\exp \left[(\frac{4}{\Ttot^2}-1)
\int_{\omega-\frac{\alpha_w\Ttot^2}{2\pi} \tau}^{\omega}d\omega'\tilde P(\omega')
\right] \ea If we focus on the most singular part as $x$ becomes small,
then $P(z)=\frac{1}{z}$ and $\tilde P(\omega)=\frac{1}{\omega}$ so that
\be\label{gen_sol} x f^{(T)}(x,\tau)=
e^{-\frac{\alpha_w\Ttot^2}{2\pi}\tau\log \frac{1}{x}}
\;\int_{c-i\infty}^{c+i\infty}\frac{d\omega}{2\pi i}\; x^{-\omega}
\;\tilde{f}_0^{(T)}(\omega)\;\;\left(\frac{\omega}{\omega-
\frac{\alpha_w\Ttot^2}{2\pi}\tau}\right)^{(\frac{4}{\Ttot^2}-1)} \ee

Let us first recover the case $T=1$, i.e. $\Ttot^2=2$. Then the solution
is: \be\label{DL1} x f^{(1)}(x,\tau)= e^{-\frac{\alpha_w}{\pi}\tau\log
\frac{1}{x}} \;\int_{c-i\infty}^{c+i\infty} \frac{d\omega}{2\pi
i}\;x^{-\omega} \;\tilde{f}_0^{(1)}(\omega)\;\;\left(\frac{\omega}{\omega-
\frac{\alpha_w}{\pi}\tau}\right),~~~~~~(T=1) \ee Note the pole at
$\omega=\alpha_w\tau/\pi$ which, in the small-$x$ region, implies the {\it
cancellation} of the form factor in front for any initial condition
$\tilde{f}_0^{(1)}(\omega)$. In particular, if we consider a flat distribution at
$\tau=0$, $xf_0^{(1)}(x)=F_1\Rightarrow
\tilde{f}_0^{(1)}(\omega)=\frac{F_1}{\omega}$, then $xf^{(1)}(x,\tau)=F_1=const$
also, as in (\ref{T_1}), so that no large terms proportional to $\log x$
are generated at all.

\begin{figure}
      \centering \includegraphics[height=50mm] {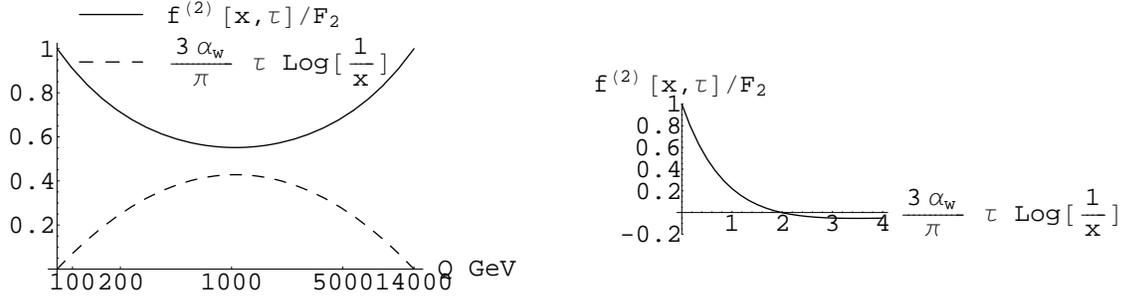}
      \caption{\label{fig4} Plots of
$\frac{3\alpha_W}{\pi}\tau\log\frac{1}{x}$ and of the structure function
$xf^{(2)}(x,\tau)/F_2$ in eq (\ref{tt2}) as functions of $Q$ for $x=
Q/\sqrt{s}$ and $\sqrt s=$14 TeV, and of $xf^{(2)}(x,\tau)/F_2$ in eq
(\ref{tt2}) as function of $\frac{3\alpha_W}{\pi}\tau\log\frac{1}{x}$ .  }
         \end{figure}

The case $T=2$ is more involved, but is still manageable analytically for
 $\tilde{f}_0^{(2)}(\omega)=F_2/\omega$, corresponding to a flat initial
 distribution. In such case the $\omega$-integral in (\ref{gen_sol}) can be
 expressed in terms of a confluent hypergeometric function as follows
 \ba\label{tt2} x f^{(2)}(x,\tau)&=&~ e^{-\frac{3\alpha_w}{\pi}\tau\log
 \frac{1}{x}}\; \int_{c-i\infty}^{c+i\infty}\frac{d\omega}{2\pi i}\;
 x^{-\omega}\;\; \frac{F_2}{\omega}\;\;\left(\frac{\omega-
 \frac{\alpha_w}{\pi}\tau}{\omega}\right)^{\frac{1}{3}}\\ \nonumber
 &=&~F_2~ e^{-\frac{3\alpha_w}{\pi}\tau\log\frac{1}{x}}\;\;F[-\frac13, 1,
 \frac{3\,\alpha_w}
 {\pi}\tau\log\frac{1}{x}]\simeq-\frac{F_2}{3\Gamma(2/3)}\;\;\left(\frac{3\,\alpha_w}{\pi}\,t\;\log\frac{1}{x}\right)^{-4/3}~~~~(T=2)
 \ea where the last behaviour holds for $\alpha_w\tau\log 1/x\gg 1$.

\vspace{0.3cm}

Let us now try to estimate the size of the effects we have calculated.
At LHC, if we consider $x_1$ and $x_2$ to be of the same order,
 we can approximate $x_1\sim x_2\equiv x= Q/(14000\, GeV)$; 
then the variable
  $\frac{3\,\alpha_w}{\pi}\tau\log 1/x$ has a maximum for
  $Q^2=M\sqrt{s}\simeq 1$ TeV and  takes the maximum value
  $\frac{3\,\alpha_w}{8\,\pi}\log^2\frac{s}{M^2}\simeq 0.43 $ corresponding
  to a depletion factor for eq (\ref{tt2}) of $\sim 0.57 \;F_2$.  If we
  include the exponential factor $\exp\left[-3\frac{\alpha_w}{4 \pi}
  \log^2\frac{Q^2}{M^2} \right]$ coming from the factorization of the
  overlap matrix (\ref{fact}), we reach a depletion of $\sim 0.46 \;F_2$
  that corresponds to corrections at the 50 $\%$ level ! 
The double-log dependence is  nontrivial (see Fig.4):
starting from the naive form factor, it changes sign, eventually, for large
(unrealistic) values of $\alpha_W\tau\log 1/x$.  The reason for that is
that the virtual and real emission flavour factors are of opposite signs.

We have so far considered initial conditions which allow a simple
 analytical understanding of the solution.  In a realistic case, one should
 set up the appropriate initial conditions by projecting out the various
 $T$-dependent components of the overlap matrix occurring in
 eq. (\ref{fact}) according to the general formula \be\label{expansion}
 \ov^H=\sum_{t_1,t_2\dots t_n} O^H_{t_1t_2\dots t_n}{\cal{P}}_{t_1t_2\dots
 t_n} \ee where $\ov,{\cal{P}}_{t_1t_2\dots t_n}$ are operators acting on
 the $n$ external legs indices, and $O_{t_1t_2\dots t_n}$ are the
 coefficients of the expansion. The projectors satisfy, by definition:
 \be\label{prop} \bom{T}_j^2{\cal{P}}_{t_1t_2\dots
 t_n}=t_j(t_j+1){\cal{P}}_{t_1t_2\dots t_n},j=1\dots n \qquad
 \bom{T}_{tot}{\cal{P}}_{t_1t_2\dots t_n}=0 \ee and have been constructed
 in various cases in refs \cite{CCC01}.

\section{Small-$x$ evolution in BFKL-type approach}

The BFKL approach \cite{BFKL} was originally proposed for massive vector
 bosons, and has been recently revisited, and applied to electroweak theory
 and to its symmetry breaking in~\cite{BFKLew}. Here we work in the $s\gg
 Q^2\gg M^2$ regime where global-symmetry restoration is expected and we
 take a simplified approach, in which all vector bosons have the same mass,
 which acts as symmetry breaking scale and as infrared cutoff. With such
 simplifications, and using the notation $Y=log \frac{1}{x}$ and $t=log
 \frac{\kv^2}{M^2}$, the weak isospin BFKL equation can be written in the
 following form \ba\label{BFKL} \frac{\partial}{\partial Y}{\cal
 F}^{(T)}(t,Y)=-\frac{\alpha_w}{\pi}\; \frac{\Ttot^2}{2}\;t\;{\cal
 F}^{(T)}(t,Y) +\frac{\alpha_w}{\pi}\; (C_A-\frac{\Ttot ^2}{2})\;\;\int
 \frac{d^2\kv'}{\pi}\;{\bf K}(\kv ,\kv')\; {\cal F}^{(T)}(t',Y) \ea where
 $t'=log\frac{\kv^{'2}}{M^2}$.  We note that the diagonal term in the
 r.h.s. is proportional to the vector boson reggeon intercept
 $\omega_V(\kv^2)=-(\alpha_w/\pi)\log \kv^2/M^2$, and becomes identical to
 it for $T=1, \Tv^2=2$. Furthermore, the flavour factors are, once again,
 the same as in the eikonal and collinear evolution equation (\ref{eqf})
 analyzed before.

Finally, the (regularized) kernel ${\bf K}$ has the spectral representation
\be {\bf K}(\kv',\kv)=\frac{1}{\kv^2}\int \frac{d\gamma'}{2
\pi\,i}\;\chi(\gamma')\; \left(\frac{\kv^{2}}{\kv^{'2}}\right)^{\gamma'}
\ee where $ {\chi}(\gamma)=\frac{1}{\gamma}+2\,
\psi(1)+\psi(1+\gamma)-\psi(1-\gamma)\sim \frac{1}{\gamma}+0(\gamma^2)$ 
(with $\psi$ the digamma function) is
the BFKL eigenvalue function , according to the equation \be \int \frac{d
\kv^{'\,2}}{\pi} \;\kv^{'\,2\,(\gamma-1)}\;{\bf K}(\kv,\kv')=
\;\chi(\gamma)\;\kv^{2\,(\gamma-1)} \ee

It is then convenient to introduce the $\gamma$-representation ($Y
\equiv\log\frac1x$) \ba k^2\f(k^2, Y)=\int\frac{d\gamma}{2\pi i}e^{\gamma
t}\tilde{\f}(\gamma, Y);~~~~
\tilde{\f}(\gamma,Y)=\int_0^{\infty}dk^2e^{-\gamma t}\f(k^2, Y) \ea and to
rewrite eq.~(\ref{BFKL}) as a differential equation \be\label{ga_ev}
\frac{\partial}{\partial Y}\tilde{\f}^{(T)}(\gamma,
Y)=\frac{\alpha_w}{\pi}\; \frac{\Ttot^2}{2}\;\;
\frac{\partial}{\partial\gamma}\tilde{\f}^{(T)}(\gamma,
Y)+\frac{\alpha_w}{\pi}\; (C_A-\frac{\Ttot
^2}{2})\;\;\chi(\gamma)\;\tilde{\f}^{(T)}(\gamma, Y) \ee This equation is
now of the same form as eq.~(\ref{om_ev}), with the variables $\tau,
\omega$ interchanged with $Y, \gamma$ and, by the same manipulations,
admits the general solution \ba\label{gensolga}
k^2\f^{(T)}(k^2,Y)\equiv\f^{(T)} (t, Y)&=&
\int_{c-i\infty}^{c+i\infty}\frac{d\gamma}{2\pi i}\; e^{\gamma t}\;
\tilde{\f_0}^{(T)}(\gamma+\frac{\alpha_w\Ttot^2}{2\pi}Y)\;\;\exp\left[(\frac{4}{\Ttot^2}-1)
\int_\gamma^{\gamma+ \frac{\alpha_w\Ttot^2}{2\pi}Y}d\gamma' \chi(\gamma')
\right] \\ &=&e^{-\frac{\alpha_w\Ttot^2}{2\pi}t Y}\;
\int_{c-i\infty}^{c+i\infty}\frac{d\gamma}{2\pi i}\; e^{\gamma t}\;
\tilde{\f}_0^{(T)}(\gamma)\;\;\exp \left[(\frac{4}{\Ttot^2}-1)
\int_{\gamma-\frac{\alpha_w\Ttot^2}{2\pi} Y}^{\gamma}d\gamma'\chi(\gamma')
\right] \ea

Such expressions look very similar to the general solution for the
DGLAP-type density $f(x, \tau)$, with the crucial difference that the
initial condition is now set at $Y=0$ instead of $\tau=0$. This means that,
in order to relate the two kinds of densities one should consistently
relate the boundary conditions too.  In particular, in the collinear limit
for which $\chi(\gamma)\simeq\frac{1}{\gamma}$, we obtain
\be\label{gen_sol_ga} {\cal F}^{(T)}(t,Y)=
e^{-\frac{\alpha_w}{\pi}\;\frac{\Ttot^2}{2}\,Y\;t}\;\; \int\frac{d
\gamma}{2 \pi \,i}\;\tilde{ \cal F}^{(T)}(Y=0, \gamma)\;\;e^{\gamma\;t}\;
\left( \frac{\gamma}{\gamma-\frac{\alpha_w}{\pi}\frac{\Ttot^2}{2}\;Y}
\right)^{\frac{4}{\Ttot^2}-1} \ee which will now be related to the solution
(\ref{gen_sol}) in the DGLAP approach by a proper choice of initial
condition.

\subsection{Solutions for $T=0$ and $T=1$}

The $T=0$ equation in (\ref{gen_sol_ga}) is QCD-like, and reads

\be\label{bfklT_0} \f^{(0)}(t,Y)=\frac{1}{2\pi
i}\;\int_{c-i\infty}^{c+i\infty}d\gamma\;\; \exp\left[\gamma
t+\frac{\alpha_W}{\pi}\frac{C_A}{\gamma}Y\right]\;\;
\tilde{\f}_0^{(0)}(\gamma),~~~(T=0) \ee to be compared to the DGLAP-type
solution (\ref{T_0}).  The initial condition corresponding to
$f_0(\omega)=F_0/\omega$ -- a constant in $Y$ space -- turns out to be
simply $\tilde{\f}_0(\gamma)=F_0$ -- a delta-function in $t$ space.

The corresponding saddle point estimates are, according to eq.~(\ref{DL}),
\ba x f^{(0)}(x, t)&\simeq& F_0 \;\;\left(
4\pi\sqrt{\frac{\alpha_w}{\pi}C_A t Y}\right)^{-1/2}
\;\;\exp\left[2\sqrt{\frac{\alpha_w}{\pi}C_A t Y }\right]\\ \nonumber k^2
\f ^{(0)}(k^2, Y)&\simeq& \left(\frac{\alpha_w C_A Y}{\pi t}\right)^{1/2}\;
x f^{(0)}(x, t)\; \simeq\; \frac{\partial xf^{(0)}(x,t)}{\partial
t}~~~~~(T=0) \ea thus justifying the customary name of ``unintegrated PDF''
for $\f (t, Y)$ in this case.

The $T=1$ case is again simplified by the presence of the simple pole at
$\gamma=\alpha_w Y/\pi$ in (\ref{gen_sol_ga}). By taking the initial
condition $\tilde{\f}_0^{(1)}(\gamma)=F_1$, corresponding to a delta
function in $t$-space, we get the solution \be\label{T1_Y} k^2 \f
^{(1)}(k^2, Y)=F_1 (\delta(t)+\frac{\alpha_w}{\pi} Y \theta(t))~~~~~(T=1)
\ee which can be easily double-checked by using the collinear approximation
$K\simeq\Theta(k^2-k'^2)/k^2$ for the BFKL kernel.

Let us remark that the solution (\ref{T1_Y}) is not the only one without
double-logs. From eq.(\ref{gensolga}) we obtain a particular
$Y$-independent solution \be
\f^{(1)}(t,Y)=\f_0^{(1)}(t)=\int_{c-i\infty}^{c+i\infty}\frac{d\gamma}{2\pi
i} e^{\gamma t}\exp[-\int^{\gamma}d\gamma'\chi(\gamma')] \ee which, in the
collinear approximation $\chi(\gamma)\simeq 1/\gamma$, yields just
$\f^{(1)}(t,Y)=$const.  This kind of solution corresponds, in our
simplified approach, to the gauge-boson Regge pole (having unit intercept)
and realizes, therefore, the so-called bootstrap of the adjoint
representation \cite{BFKL,boot,BFKLew}.

\subsection{Solutions for generic $T$ values }

The previous examples suggest to look for a general relation between
BFKL-type and DGLAP-type densities by assuming the related initial
conditions \ba x
f^{(T)}(x,t=0)&=&F_T,\;\;~~~f_0^{(T)}(\omega)=F_T/\omega,~~~~\\ \nonumber
\f^{(T)}(t,Y=0)&=&F_T\;\delta(t),~~~\tilde{\f}_0^{(T)}(\gamma)=F_T \ea In
fact, we notice that the expressions in eq.(\ref{gen_sol})
(eq.~(\ref{gen_sol_ga})) can be almost identified by the rescaling
$\omega\rightarrow\alpha_w\Tv^2\tau/\pi ~\eta$
($\gamma\rightarrow\alpha_w\Tv^2 Y/\pi ~\eta$), which singles out the
double-log variable $\alpha_w\Tv^2 \tau Y/\pi$. The two kinds of densities
are thus simply related, except for a $\gamma$-integral Jacobian factor
which is compensated by a $t$-derivative in the $\omega$-integral as
follows: \be e^{\frac{\alpha}{\pi}\frac{\Ttot^2}{2}\,Y\;t}~{\cal
F}^{(T)}(t,Y)= \frac{\partial}{\partial
t}\left(e^{\frac{\alpha}{\pi}\frac{\Ttot^2}{2}\,Y\;t} \;\;x
f^{(T)}(x,t)\right) \ee This equation extends to generic $T$ values the
identification of ${\cal F}^{(T)}(t,Y)$ as a sort of ``unintegrated''
density, compared to the integrated distribution function $xf^{(T)}(x,t)$.

\section{Conclusions}

We have investigated the structure of enhanced EW corrections to a basic
Drell-Yan-type inclusive process (like $WW(s)\rightarrow$ jet$(Q)+X$) in
the kinematical limit where $x\equiv x_1x_2=Q^2/s\ll 1$.  This regime is
characterized by three different scales $s\gg Q^2\gg M^2$ and the gauge
boson emission generates several large logarithms, of high-energy type
$\sim\log \frac{s}{Q^2}\sim\log (x_1x_2)$ and of infrared or collinear type
$\sim t\equiv\log \frac{Q^2}{M^2}$.  Due to its nonabelian nature, the
eikonal $W$-emission (sec. 3) naively predicts the presence of various
kinds of uncanceled double log corrections, $\log^2 x_i, \quad \log
x_i\;\log \frac{Q^2}{M^2}$ and $\log^2 \frac{Q^2}{M^2}$, arising in the
eikonal exponent, of type $\log^2 \frac{s}{M^2}$. We have first justified a
factorized structure of the cross-section, in which the double-log form
factor occurs at scale $Q^2$, while the ``incoming parton'' distribution
functions (which also involve leptons and gauge bosons) only have collinear
and high-energy logs.

Then, by solving both the EW collinear evolution equations\cite{CCC01} and
the EW BFKL dynamics\cite{BFKLew} (secs. 4 and 5), we have explicitly
computed the dependence of the PDFs on such enhanced variables, and we find
peculiar features, depending on the values of the total $t$-channel weak
isospin $\Ttot^2=T(T+1)$:

\begin{itemize}

\item For $T=0$ the EW corrections have the same structure as the QCD ones
(\ref{DL}) and thus show a customary double-log enhancement.
\item For $T=1$, potential large and negative $\alpha_w t \log \frac{1}{x}$
 corrections can appear but, due to the fact that the Casimir charges for
 real and virtual $W$-emission are equal, a cancellation mechanism is at
 work (\ref{DL1}), leaving only the exponential form factor ($\exp \left[
 -\frac{\alpha_w} {2 \pi} \log^2\frac{Q^2}{M^2}\right]$) already
 incorporated in the factorization formula for the overlap matrix, eq
 (\ref{fact}).
\item For $T=2$ the Casimir charges for real and virtual $W$-emission are
 different (and of opposite sign) so the previous mechanism of cancellation
 fails and large $\alpha_w t \log \frac{1}{x}$ and non-trivial corrections
 to the form factor at scale $Q^2$ can be present at LHC also, as shown in
 Fig. 4.

\end{itemize}

The above analysis tells us  that, with relatively low $Q^2$, large EW
  corrections can be present only for cross sections initiated by two
  transverse gauge bosons, being the only partons supporting the $T=2$
  total t-channel isospin component.  For instance, in ref \cite{generic}
  we have already investigated the isospin decomposition of the partonic
  cross section $WW\rightarrow f \bar f$, inclusive over the final
  fermions, at the double log level. 
  Clearly a detailed
  phenomenological investigation at the double and single log level
  proposed here has to be performed for various specific
  processes.

Other possible large effects coming from the small-$x$ EW corrections are
 in cross sections with more than two detected isospin charged legs (that
 can be both in the initial or in the final state).  In this case,
 depending on type of measurement, the total isospin of the overlap can be
 $T\geq 2$, with nontrivial double-log effects.  Again, explicit
 phenomenological applications have to be investigated.

\vspace{0.5cm}

{\bf \large Acknowledgments:} 
 M.C. and P.C. wish to thank the CERN Theory Division for hospitality while
part of this work was being done.
Work supported in part by a PRIN grant (MIUR, Italy).


\begin{thebibliography}{99}

\bibitem{sirlin}


M.~Kuroda, G.~Moultaka and D.~Schildknecht,
Nucl.\ Phys.\ B {\bf 350} (1991) 25;

G.~Degrassi and A.~Sirlin,
Phys.\ Rev.\ D {\bf 46}, 3104 (1992);

A.~Denner, S.~Dittmaier and R.~Schuster,
Nucl.\ Phys.\ B {\bf 452}, 80 (1995);

A.~Denner, S.~Dittmaier and T.~Hahn,
Phys.\ Rev.\ D {\bf 56}, 117 (1997),
Nucl.\ Phys.\ B {\bf 525}, 27 (1998);

W.~Beenakker, A.~Denner, S.~Dittmaier, R.~Mertig and T.~Sack,
Nucl.\ Phys.\ B {\bf 410}, 245 (1993);

W.~Beenakker, A.~Denner, S.~Dittmaier and R.~Mertig,
Phys.\ Lett.\ B {\bf 317}, 622 (1993).

M.~Beccaria, G.~Montagna, F.~Piccinini, F.~M.~Renard and C.~Verzegnassi,


\bibitem{CC} P.~Ciafaloni and D.~Comelli,
Phys.\ Lett.\ B {\bf 446}, 278 (1999).

\bibitem{FadinCC} V.~S.~Fadin, L.~N.~Lipatov, A.~D.~Martin and M.~Melles,
Phys.\ Rev.\ D {\bf 61} (2000) 094002;

P.~Ciafaloni, D.~Comelli,
Phys.\ Lett.\ B {\bf 476} (2000) 49.

J.~H.~Kuhn, A.~A.~Penin and V.~A.~Smirnov,
Eur.\ Phys.\ J.\ C {\bf 17}, 97 (2000);

J.~H.~Kuhn, S.~Moch, A.~A.~Penin, V.~A.~Smirnov,
Nucl.\ Phys.\ B {\bf 616}, 286 (2001) [Erratum-ibid.\ B {\bf 648}, 455
(2003)];

M.~Melles,
Phys.\ Rept.\ {\bf 375}, 219 (2003);


J.~y.~Chiu, F.~Golf, R.~Kelley and A.~V.~Manohar, 
  arXiv:0712.0396 [hep-ph].


\bibitem{CCC00} M.~Ciafaloni, P.~Ciafaloni and D.~Comelli, Phys.\ Rev.\
Lett.\ {\bf 84}, 4810 (2000); Phys.\ Lett.\ B {\bf 501}, 216 (2001);
Nucl.Phys. B 589 359 (2000); Nucl.Phys. {\bf B613 }, 382 (2001);
  Phys.\ Rev.\ Lett.\ {\bf 87} (2001) 211802

P.~Ciafaloni, D.~Comelli and A.~Vergine, JHEP {\bf 0407}, 039 (2004).

M.~Ciafaloni, 
arXiv:hep-ph/0612067.


\bibitem{generic} M.~Ciafaloni, P.~Ciafaloni and D.~Comelli, Phys.\ Lett.\
  B {\bf 501}, 216 (2001);

\bibitem{CC06} P.~Ciafaloni and D.~Comelli, JHEP {\bf 0609}, 055 (2006).

 \bibitem{Baur} U.~Baur, Phys.\ Rev.\ D {\bf 75} (2007) 013005;

 R.~S.~Thorne, 
arXiv:0711.2986 [hep-ph].  

\bibitem{excl}
See S.~Pozzorini, ``Electroweak Radiative Corrections At High Energies,''
arXiv:hep-ph/0201077 and references therein.


 \bibitem{2loop} A Denner, B. Jantzen and S. Pozzorini Nucl. Phys B {\bf
761} (2007) 1  and arXiv:0801.2647.



\bibitem{CCC01} M.~Ciafaloni, P.~Ciafaloni and D.~Comelli, Phys.\ Rev.\
 Lett.\ {\bf 88}, 102001 (2002);

P.~Ciafaloni and D.~Comelli, 
JHEP  {\bf 0511} (2005) 022.

\bibitem{DGLAP}

V. Gribov, L. N. Lipatov, Sov. J. Nucl. Phys {\bf 15} 438 (1972);

L. N. Lipatov, Sov. J. Nucl. Phys {\bf 20}, 94 (1972);

G. Altarelli, G. Parisi, Nucl. Phys. {\bf B 126}, 298 (1977);

Y. Dokshitzer, Sov. Phys. JETP {\bf 46}, 641 (1977).

\bibitem{BFKL} L.N.~Lipatov, Sov. J. Nucl. Phys {\bf 23}, 338 (1976);

E. A.~ Kuraev, L.N.~Lipatov and V. S. Fadin, Sov. Phys. JETP {\bf 45}, 199
(1977);

Y. Y. Balitskii and L. N. Lipatov, Sov. J. Nucl. Phys {\bf 28}, 822 (1978).


\bibitem{BFKLew}
%
J.~Bartels, L.~N.~Lipatov and K.~Peters,
Nucl.\ Phys.\ B {\bf
  772}, 103 (2007).

\bibitem{SCB}
S.~Catani and M.~Ciafaloni, Nucl.\ Phys.\ B {\bf 249} (1985) 301;

G.~Sterman,  Nucl.\ Phys.\ B {\bf 281} (1987) 310; J.~Botts and
  G.~Sterman, Nucl.\ Phys.\ B {\bf 325} (1989);

R.~Bonciani, S.~Catani, M.~L.~Mangano and P.~Nason, Phys.\ Lett.\ B {\bf 575} (2003)
  268.

\bibitem{DM}

Yu.~L.~Dokshitzer and G.~Marchesini,   JHEP {\bf 0601} (2006) 007.
\bibitem{BCM} A.~Bassetto, M.~Ciafaloni and G.~Marchesini, Phys.\ Rep.\
{\bf 100} (1983) 201.

\bibitem{boot}

V. S. Fadin and R.~Fiore, Phys.\ Lett.\ B {\bf 440} (1998) 359.


 \end{thebibliography}
\end{document}